\documentclass[aps,preprint,groupedaddress,showpacs,amsmath,amssymb,endfloats*]{revtex4}
\usepackage{graphicx}
\usepackage{bm}

\begin{document}

\title{Collisional decoherence in trapped atom interferometers that use non-degenerate sources}
\author{James A. Stickney}
 \affiliation{Space Dynamics Laboratory, Bedford, MA 01730, USA}
\email[]{AFRL.RVB.PA@hanscom.af.mil}

\author{Matthew B. Squires}
\affiliation{Air Force Research Laboratory, Hanscom AFB, MA 01731, USA}

\author{James Scoville}
\affiliation{Air Force Research Laboratory, Hanscom AFB, MA 01731, USA}

\author{Paul Baker}
\affiliation{Air Force Research Laboratory, Hanscom AFB, MA 01731, USA}

\author{Steven Miller}
\affiliation{Air Force Research Laboratory, Hanscom AFB, MA 01731, USA}

\begin{abstract}
The coherence time, and thus sensitivity, of trapped atom
interferometers  that use non-degenerate gases are limited by the
collisions between the atoms.  An analytic model that describes the
effects of collisions between atoms in an interferometer is
developed.  It is then applied to an interferometer using a harmonically trapped non-degenerate
atomic gas that is manipulated with a single set of standing wave laser
pulses.  The model is used to find the optimal operating conditions of
the interferometer and direct
Monte-Carlo simulation of the interferometer is used to verify the
analytic model.      
\end{abstract}

\maketitle

\section{Introduction}\label{sec:introduction}

To date cold atom interferometers have demonstrated rotation
sensitivities comparable to ring laser and mechanical gyroscopes
\cite{durfee06}.  Several atom interferometer schemes have been
realized, and thus far free space fountain and beam configurations, that utilize light
pulses to manipulate the atoms, have demonstrated the greatest
sensitivities \cite{durfee06, peters99, Hogan08}.  
While the lack of external
potential reduces systematic errors,  atom interferometry in free
space is limited by acceleration due to gravity.  In particular,
the precision of an interferometer is directly proportional to the
interrogation time and in free space this time is limited by the size of the vacuum
chamber.  In the most sensitive free space atom interferometers, the atomic clouds
travel up to 10 meters \cite{Hogan08}.  The large scale of free space 
interferometers limits their applications.

There is currently a great effort being made to reduce
the size of atom interferometers while simultaneously increasing their
sensitivity.  One straightforward way to achieve 
this goal is to develop interferometers that trap the atoms in an 
external potential for the duration of the interferometer cycle.
The external potential prevents the atoms from falling due to gravity,
and keeps the atomic gas from expanding in the vacuum chamber.
As a result, the interferometer cycle time is not as limited by the size
of the chamber.

Several groups have built trapped atom interferometers using atomic
gases that are both above and
below the recoil temperature \cite{Schumm05,Wang05,Garcia06,Horikoshi06,Jo07, Wu07}.  
To date, all interferomters that use gases below the recoil
temperature have utilized atoms in a nearly pure Bose-Einstein
condensate (BEC).  

If the gas is cooled below the recoil temperature and is split using
a laser pulse, a large relative separation between each arm of
the interferometer can be achieved \cite{Garcia06}.  
By exposing the atoms in each arm to a different environment, precision 
measurements of localized phenomena can be performed.  For example, the AC stark shift 
in $^{87}\mbox{Rb}$ was recently measured  by exposing the atoms 
in one arm of an interferometer to laser light \cite{Deissler08}.

In some applications, such as the sensing of rotations and accelerations, cloud separation 
is not necessary and interferomters that use non-degenerate source are
sufficient.  These sources can be produced by laser cooling alone.    
Additionally, 
non-degenerate atomic gases have a much lower density compared to a BEC 
and therefore experience a weaker mean-field
potential.  The mean-field potential directly couples number uncertainty into
phase uncertainty via number dependent phase diffusion \cite{Jo07,
  javanainen97, leggett98, javanainen98}.
One advantage of working with laser cooled gases is that dephasing 
due to number fluctuations is ameliorated. 

Besides the elimination of mean-field effects, atom
interferometers that use laser cooled atomic
gases are less sensitive to heating due to imperfections in
the confining potential than a  BEC.  Atoms in a pure BEC experience no
momentum changing collisions with other atoms in the same mode.  Therefore, an
interferometer that uses a pure BEC will experience little 
decoherence due to collisions.   However, if a BEC is heated,
atoms will leave the condensate and will experience an increase in the
collision frequency.  As a
result, the decoherence rate will increase with
temperature.  On the other hand, if the interferometer uses a laser
cooled gas with a temperature much greater than the BEC transition
temperature, the density will decrease as the temperature increases and the
decoherence rate will decrease if the gas is inadvertently heated.

Several different methods for building atom interferometers using
laser cooled gases have been developed \cite{cronin07} and 
time-domain atom
interferometers that use a single internal quantum state \cite{cahn97}
lend themselves naturally for use with trapped atomic gases.  
This type of interferometer uses a series of optical standing waves to
manipulate the external states of the atoms in the cloud.  The
interferometric cycle begins by loading an atomic cloud in a magneto-optical trap.  The
trap is switched off and the atomic cloud begins to fall due to
gravity.  At the time $t = 0$, the cloud is illuminated with a
short pulse from the standing wave laser field.  
Shortly after the pulse, the cloud has a density modulation with a
period of $\lambda /2$, where $\lambda$ is the wavelength of the laser
field.  The density modulation then disappears because of the thermal
motion of the atoms in the cloud.  At the time $t = T$ the gas is
illuminated with a second pulse.  Due to the Talbot-Lau effect, there
is an echo of the density modulation at the times $t = n T$, for integers
$n \ge 2$.  If the atomic cloud experiences a non-uniform potential
during the interferometer cycle, the density echos will be shifted
relative to the initial modulation.  The shift in the phase of the
modulation can be determined by reflecting a
probe pulse from a single laser beam off of the 
echo.  The phase of the reflected probe pulse is directly proportional
to the phase shift in the density modulation.  By interfering the reflected probe
pulse with a reference beam, the interferometer's signal can be read.

Recently, a trapped time domain atom interferometer was built by the group
at Harvard \cite{Wu07}.  This interferometer used an atom wave guide to confine the atoms
in the perpendicular directions while allowing them to freely propagate
along the parallel direction.  A series of standing wave laser pulses were applied to the
atoms, such that the wave vectors of the lasers pointed along the free
direction of the guide.  The Harvard group demonstrated that it
is possible to electronically move the wave guide back and fourth
perpendicular to the free direction of the wave guide so that the arms of the
interferometer enclose an area, making the interferometer
 sensitive to rotations \cite{Wu07}.

A major difficulty with all trapped atom interferometers that use
optical pulses is that the residual
potential along the guide causes decoherence \cite{horikoshi07,
Wu07_2, burke08, stickney08}.  The groups that have built BEC based
interferometers have mitigated the decoherence by either using a
double reflection geometry or using the classical turning points of
the residual potential to reflect the atoms.  
The Harvard group has reduced the effects of the residual potential by
using an interferometric cycles with several laser pulses \cite{su08}.  Although 
this multi-pulse scheme greatly increases the coherence time of the 
interferometer, it also reduces the number of atoms participating 
as well as reducing the area inclosed by the interferometer.

\begin{figure}
\includegraphics[width=8.6cm]{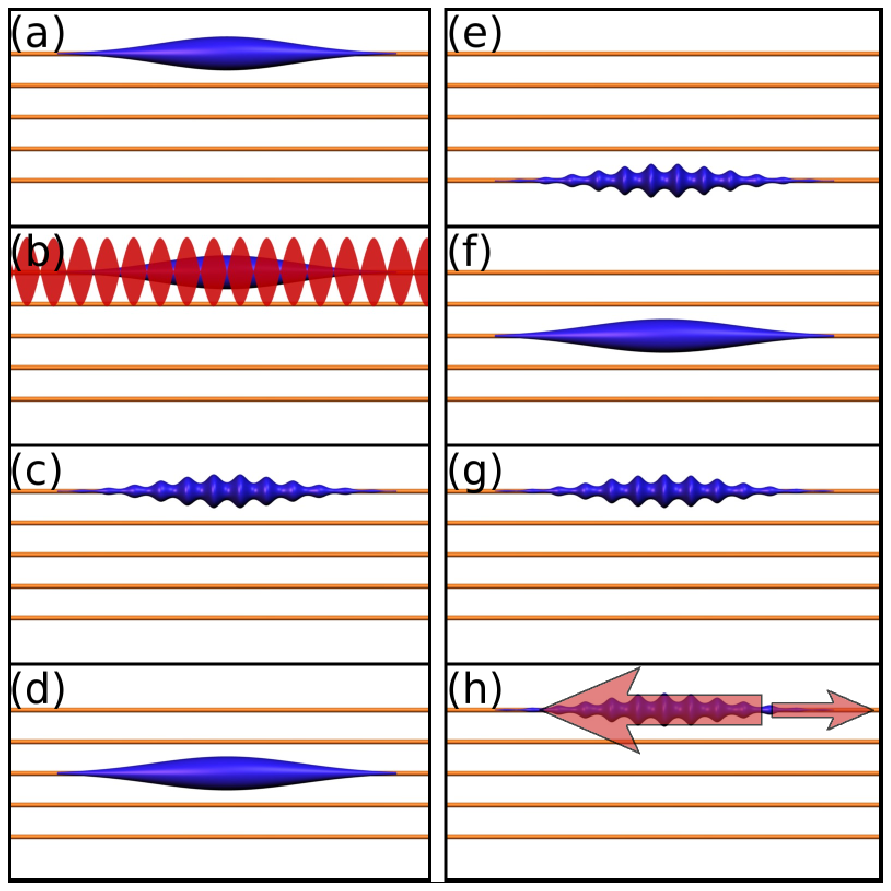}%
\caption{ \label{fig:cartoon}
(color online) A schematic of the interferometer cycle.  (a) The cold
gas is loaded into a cigar shaped trap, created with the upper most
horizontal wire  (b) The atomic cloud is illuminated with a
standing wave laser field.   (c) After the laser pulse, a
density modulation appears across the cloud.  (d) The density
modulation disappears and the trap is moved
downwards by cycling the current in the wires. (e) The trap is above the bottom
wire at a half period of the parallel trap and the density modulation
reappears across the cloud.  (f) The trap is moved upward and the
density modulation disappears.  (g) The
trap reaches the top wire at one trap period of
the parallel trap  and the density modulation reappears for a second
time.  
(h)  A probe pulse is reflected off of the cloud.  
}
\end{figure}

We are currently developing a trapped atom gyroscope that uses
a laser cooled atomic gas and  avoids
decoherence due to the residual potential by using
classical turning points to reflect the atoms.  
Rather than utilizing the Talbot-Lau effect, the density modulation
will echo twice every oscillation of the atoms in the parallel
direction. 
Figure \ref{fig:cartoon} is a schematic of the interferometer cycle.
(a) Initially, a laser cooled atomic gas is loaded into a cigar shaped trap.  The trap in the perpendicular
direction is created with the upper most horizontal wire plus a
uniform bias field.  The relatively weak trap in the parallel
direction is created using vertical wires that are not shown.  (b) At the beginning
of the interferometer cycle $t=0$, the atomic cloud is illuminated with a
standing wave laser field. The atoms are accelerated towards the nodes
of the laser field.  (c) Immediately after the laser pulse, the atoms
move towards the location of the nodes and density modulation appears 
across the cloud.  (d) The density modulation disappears due to the 
thermal motion of the atoms.  Simultaneously the trap is moved
downwards by cycling the current in the wires.  If the interferometer
is rotating about the plane of the paper with frequency $\Omega$, the Coriolis force will
accelerate the cloud in the parallel direction.  (e)  The cycling of the
currents in the wires is timed so that the trap is above the bottom
wire at half a period of the parallel trap $t=T/2$, where $T$ is the
trap period in the parallel direction.  Near $t=T/2$
there in an echo of the density modulation
across the cloud.  (f) The trap is moved upward.  The
Coriolis force decelerates the cloud resulting in a displacement
of the cloud that is directly proportional to the rotation frequency
of the interferometer.  (g) At the time $t=T$, the trap returns to the 
top wire and there is a second echo of the  density modulation.  The
modulation is shifted due to the rotation of the interferometer.  The
cycle (c) through (g) can be repeated many times.  Since the
oscillating Coriolis force is resonant with the parallel trap frequency, 
the shift in the displacement of the cloud will increase after each cycle. 
(h)  After $n$ cycles, the displacement of the cloud is precisely
measured by reflecting a probe beam off of the density modulation and
interfering the reflected light with a reference beam.   

The probe pulse only interacts strongly with the cloud
when there is a density modulation across the cloud.  As a result, the
probe pulse can be  longer than the duration of the modulation echo.  
Small fluctuations in the trap frequency can be measured
simultaneously with the phase shift and using thermal atoms avoids
the critical timing needed when a BEC is used  \cite{Stickney07,
  horikoshi07, burke08}.
It may prove possible to measure the interference signal more than once in any given
experiment.   As a result, it might be possible to split the atoms
once, and measure the rotation frequency several times as the cloud oscillates in the trap.

During interferometer cycle, collisions between the trapped atoms will
bring the gas back to equilibrium, causing a reduction in the
amplitude of the density modulation.  Thus, the amplitude of the
reflected probe pulse will degrade with time.  The upper limit on the
interferometer cycle time and the devices sensitivity can be
determined by analyzing the effects collisions between the atoms in
the trap.

In this paper, we present a theoretical model for our interferometer.
In Sec. \ref{sec:theory}, we present an analytic model for the
amplitude of the reflected probe pulse including the rotation of the
interferometer and the effects of
collisions between the atoms in the gas.  In Sec.~\ref{sec:discussion}
the analytic
model will be used to estimate the minimum value of the trap frequency
in the perpendicular direction, the optimal temperature of the gas, and
the optimal number of atoms to use in our upcoming
experiment and we compare the results of our analytic model with a Direct simulation
Monte-Carlo code. 
Finally Sec.~\ref{sec:conclusions} conclusions will be presented.

\section{Formulation of the problem}\label{sec:theory}

The dynamics of a dilute atomic gas above BEC 
phase transition temperature is governed by the quantum Boltzmann 
equation \cite{akhiezer81}
\begin{equation}
\frac{d}{dt}  \rho  - \frac{1}{i \hbar} [H_{eff}, \rho] =   I(\rho),\label{eq:QBE}
\end{equation}
where $\rho$ is the single particle density operator, $H_{eff}$ is the effective 
single particle Hamiltonian, and $I(\rho)$ is the collision integral.  
The effective Hamiltonian for an atom in a rotating trap and 
standing wave laser field is
\begin{equation}
H_{eff} =  \frac{p^2}{2m} + V(\pmb r) + \hbar \chi \cos(2 \pmb{k}_l \cdot \pmb r) - \pmb \Omega \cdot ( \pmb r \times \pmb p) + 2U_0 n(\pmb r) , \label{Heff}
\end{equation}
where $m$ is the atomic mass, $V$ is the external trapping potential,
 $\chi$ characterizes the strength of the  standing wave laser, $\pmb{k}_l$ is the 
wave vector of the laser field, and $\pmb
 \Omega$ is the vector 
that points along the axes of rotation with the magnitude of the
angular rotation frequency.  
The final term in Eq.~(\ref{Heff}) is the mean-field potential 
where $U_0 = 4 \pi \hbar^2 a_s / m$ characterizes strength of the atom-atom 
interactions, $a_s$ is the s-wave scattering length, 
and $n$ is the number density of the atomic gas.  Note that the mean field 
potential for a non-condensed gas is a factor of two larger than 
for a BEC with the same density.

It is convenient to recast the single particle density operator in the
Wigner function representation which is defined as
\begin{equation}
 f(\pmb r,\pmb p)  =
\frac{1}{( \pi \hbar)^3} \int d^3r'  \langle \pmb r - \pmb r' | \rho | \pmb r + \pmb r' \rangle e^{2 i \pmb {r}' \cdot \pmb p/\hbar},\label{eq:Wigner}
\end{equation}
where $|\pmb r\rangle$ are the eigenstates of the coordinate operator.  The Wigner 
function can be interpreted as the probability density of finding an 
atom at the coordinate $\pmb r$ with momentum $\pmb p$.

It will  be assumed that the standing wave laser pulse is in the Kapitza-Dirac
regime.  It is sufficiently short that 
both the free evolution of the gas and the collision integral may be 
neglected, i.e. the atoms do not move and experience no collisions
while the laser beams are on. The pulse is in this regime when 
\begin{equation}
  \label{eq:LP_INEQ}
  \tau_p \ll \frac{\lambda}{\bar v} , \tau_p \ll \frac{1}{\nu},
\end{equation}
where $\tau_p$ is the length of the pulse, $\lambda$ is the wavelength
of the laser beams, $\bar v$ is the average speed of the atoms in the
gas, and $\nu$ is the average collision frequency. 
When Eq.~(\ref{eq:LP_INEQ}) is fulfilled the dynamics of the atomic gas 
can be separated into two parts: the dynamics when the laser beams are on and
the dynamics when the laser beams are off.  

In what follows the  dimensionless coordinate $\pmb r' = 2 k_l \pmb x$, the dimensionless 
momentum $\pmb p' = \pmb p/2 \hbar k_l $ and the dimensionless time $t'
= t/t_0$ where $t_0 = m/4 \hbar k_l^2 $ will be used.  For
$^{87}\mbox{Rb}$, the characteristic time is $t_0 = 5.3~\mu
\mbox{s}$.  
Substituting Eq.~(\ref{eq:Wigner}) and (\ref{Heff}) into Eq.~(\ref{eq:QBE}) the dimensionless
equation of motion for the Wigner function $f$, when the laser beams are on, is
\begin{eqnarray}
\frac{\partial}{\partial t} f(\pmb r, \pmb p, t) &=& 
\chi \sin(2 x) \left[ f(\pmb r, \pmb p - \pmb k_l /2) -  f(\pmb r, \pmb p + \pmb k_l /2)\right],  \label{lasersON1}
\end{eqnarray}
where $\pmb k_l' = \pmb k_l / k_l$ is the direction of the standing wave laser field,
 $\chi' = t_0 \chi$ is the dimensionless laser strength, and all of the primes have been dropped.
Similarly, the dimensionless equation of motion for the Wigner function
$f$, when the laser beams are off, is
\begin{eqnarray}
\frac{\partial}{\partial t} f(\pmb r, \pmb p, t) &=& - {\pmb p}\cdot \frac{\partial f}{\partial \pmb r} + \frac{\partial V_{eff}}{\partial \pmb r} \cdot \frac{\partial f}{\partial \pmb p}  
\nonumber \\
&-&  \pmb \Omega \cdot \left(
\pmb r \times \frac{\partial f}{\partial \pmb r} + \pmb p \times \frac{\partial f}{\partial \pmb p} 
 \right) + I_{coll}, \label{BoltzGeneralRot}
\end{eqnarray}
where  $\Omega' = t_0 \Omega$, $y_0' = 2 k_l
y_0$ and once again all the 
primes have been dropped.  Our short term goal is to measure the
rotation of the Earth.  For $^{87}\mbox{Rb}$ the rotation frequency of
the Earth in our dimensionless units is $\Omega_{E} = 4 \times 10^{-10}$.
When the gas is in thermodynamic
equilibrium,  the dimensionless temperature is $T' = T/4T_R$, where $T_R =
\hbar^2 k_l^2 / m k_B$ is
the one photon recoil temperature.  For $^{87}\mbox{Rb}$, the recoil
temperature is $T_R = 350~\mbox{nK}$.

The dimensionless effective potential is
\begin{equation}\label{eq:EffectivePotential}
 V_{eff} = V + 2 g n,
\end{equation}
where $V' = t_0 V / \hbar $ is the dimensionless trapping potential, $n' = 8 k_l^3 n $ is the dimensionless
density, and $g = 8 \pi a_s k_l$ is the dimensionless mean-field
strength.  For $^{87}\mbox{Rb}$ the dimensionless mean-field strength
is $g \sim 1$.

The length scale $L$ of density changes in a magnetically trapped atomic 
gas is typically much larger than the atoms s-wave scattering length $a_s$, 
i.e. $a_s/L \ll 1$.  In this limit, the collision integral becomes independent of
the potential.  Since the atomic gas is above the BEC transition temperature,
no single quantum state has a macroscopic population and Bose enhanced scattering can
be neglected.   The dimensionless collision 
integral can be approximated with the classical collision integral
\cite{akhiezer81}
\begin{equation}
 I_{coll} = \frac{\sigma}{4 \pi } \int d^3 p_3 d\Omega |\pmb p_3 - \pmb p| \left[ f(\pmb r, \pmb p_1) f(\pmb r, \pmb p_2) - f(\pmb r, \pmb p_3) f(\pmb r, \pmb p) \right],  \label{Icoll}
\end{equation}
where $\sigma = 32 \pi a_s^2 k_l^2$ is the dimensionless collision
cross section.  For $^{87}\mbox{Rb}$ the dimensionless scattering
cross section is $\sigma = 0.2$. 

Before the laser pulse is applied, the gas is in thermodynamic 
equilibrium and it will be assumed that the laser pulse is 
sufficiently weak that the gas is always close to equilibrium.  
The Wigner function $f$ can be written as
\begin{equation}
 f = f_0 + \delta f,
\end{equation}
where $f_0$ is the equilibrium Wigner function and $\delta f$ is the
disturbance caused by the laser pulse.  When the disturbance 
is much smaller than the equilibrium $|\delta f| \ll |f|$, Eq.~(\ref{Icoll}) 
can be approximated as \cite{chapman70}
\begin{equation}
 I_{coll} =  \frac{\sigma}{4 \pi} \int d^3 p_3 d\Omega |\pmb p_3 - \pmb p| \left[ 
2 f_0(\pmb r, \pmb p_1) \delta f(\pmb r, \pmb p_2) - f_0(\pmb r, \pmb p_3) \delta f(\pmb r, \pmb p) - f_0(\pmb r, \pmb p) \delta f(\pmb r, \pmb p_3)
\right]. \label{Icoll1}
\end{equation}
Equation~(\ref{Icoll1}) is the sum of three terms, each with a simple 
physical interpretation.  The first term $2 f_0(\pmb r, \pmb p_1) \delta f(\pmb r, \pmb p_2)$ is proportional 
to the rate that an atom in $f_0$ scatters with an atom in $\delta f$ and one of the atoms scatters into the
momentum state $\pmb p$.  The second term 
 $f_0(\pmb r, \pmb p_3) \delta f(\pmb r, \pmb p)$ is proportional to
 the rate that atoms scatter out of $\delta f$ because of collisions
 with atoms in $f_0$.  The final term 
$f_0(\pmb r,\pmb p) \delta f(\pmb r, \pmb p_3)$ is the inverse of the second process. 

Only atoms in the disturbance contribute to the interference signal. 
Therefore, once an atom scatters out of the disturbance  
it no longer contributes to the interference signal.  As a result only the second term in 
Eq~(\ref{Icoll1}) contributes to the loss of the
interference signal and the collision integral Eq.~(\ref{Icoll1}) 
becomes
\begin{equation}
 I_{coll} =  \nu(\pmb r, \pmb p) (f_0(\pmb r, \pmb p) - f(\pmb r, \pmb p)),
\end{equation}
where the collision frequency $\nu$ is given by the integral
\begin{equation}
 \nu(\pmb r, \pmb p) = \sigma \int d^3 p_3 |\pmb p_3 - \pmb p| f_0(\pmb r, \pmb p_3). \label{nu1}
\end{equation}
Substituting the equilibrium distribution 
\begin{equation} \label{eq:equlib_distribution}
f_0 =\frac{1}{( 2 \pi T)^{3/2}} n(\pmb r) e^{- p^2/2 T} 
\end{equation}
into Eq.~(\ref{nu1}), the 
collision frequency Eq.~(\ref{nu1}) becomes
\begin{equation}
 \nu = \frac{2}{\pi} \rho(\pmb r)  \sigma T^{1/2} K(|\pmb p|/\sqrt{2 T}),
\end{equation}
where 
\begin{equation}
 K(\xi) = \int d\eta d\theta \eta^2 \sin \theta \sqrt{\xi^2 + \eta^2 - 2 \xi \eta \cos\theta} e^{-\eta^2}. \label{K} 
\end{equation}
The integral $K(\xi)$ can be explicitly evaluated in terms of error 
functions.  
To remove the dependnce of $\nu$ on the coordinate $\pmb r$ and
momentum $\pmb p$, Eq.~(\ref{K}) will be replaced by its value a zero
argument $K = 1$ and the density $n$ will be replaced by the averaged
density of a gas thermodynamic equilibrium in a harmonic potential.  The 
collision  frequency Eq.~(\ref{nu1}) becomes
\begin{equation}
 \nu = \frac{2^{1/2}} {(2 \pi)^2} \frac{\bar \omega^3 \sigma
   N}{T}, \label{eq:nu}
\end{equation}
where $\bar \omega = (\omega_\parallel \omega_\perp^2)^{1/3}$ is the
geometric average of the trap frequencies.  In
Sec. \ref{sec:discussion} it will be demonstrated that using
Eq.~(\ref{eq:nu}) for the collision frequency yields
accurate results when compared to a more complete description of the
atomic collisions.

For the rest of this paper, we will limit the discussion to the case of a cigar
shaped harmonic potential.   The dimensionless trapping 
potential becomes
\begin{equation}
  V = \frac{1}{2} \left\{ \omega_\parallel^2 x^2 + \omega_\perp^2
    \left[ (y-y_0)^2 + z^2  \right] \right\}, \label{eq:HarmonicPotential}
\end{equation}
where $\omega_\perp' = t_0 \omega_\perp$ and $\omega_\parallel' = t_0
\omega_\parallel$.  If the trap has frequencies $\omega_\parallel = 2
\pi \times 3~\mbox{Hz}$ and $\omega_\perp = 2 \pi \times
300~\mbox{Hz}$, 
the dimensionless trap frequencies for $^{87}\mbox{Rb}$ are $\omega_\parallel' = 10^{-4}$ and $\omega_\perp' = 10^{-2}$.

When the gas is close to thermodynamic equilibrium in a harmonic trap the
density is
\begin{equation} \label{eq:Density}
n = \frac{N \bar \omega^3}{(2 \pi T)^{3/2}} e^{- V / T },
\end{equation}
where $N$ is the number of atoms in the trap and $V$ is the potential.
Using Eq. (\ref{eq:EffectivePotential}) and (\ref{eq:Density})   
the effective potential can be expanded to fourth order as
\begin{equation}
 V_{eff} =  \left(
1 - \frac{2 g \bar \omega^3 N}{(2 \pi T)^{3/2} }
 \right) V
+ \frac{2 g \bar \omega^3 N}{(2 \pi)^{3/2} T^{7/2}} V^2 .
\end{equation}
The lowest order mean field contribution to the potential causes a small reduction of the trap 
frequency. Therefore, the oscillation period of atoms in the trap is weakly
dependent on the number of trapped atoms.  The next higher order contribution is a
weak quartic contribution to the potential.  This, and all higher order terms,
 can be neglected when
\begin{equation} \label{eq:quartic_inequality}
\frac{2 g \bar \omega^3 N}{(2 \pi)^{3/2} T^{7/2}} \ll 1. 
\end{equation}  
For example if the trap contains $7 \times 10^6$ $^{87}\mbox{Rb}$
atoms, in a trap with
frequencies  $\omega_\parallel = 2
\pi \times 3~\mbox{Hz}$ and $\omega_\perp = 2 \pi \times
300~\mbox{Hz}$, and a temperature of $40~\mu\mbox{K}$ 
($T = 30$), the left hand side of Eq.~(\ref{eq:quartic_inequality}) is about
$6 \times 10^{-8}$.  In this case, the quantric contribution to the potential can be neglected.

The analysis of the operation of the interferometer will be limited 
to the case where the splitting and read lasers beams are 
aligned with the weak axis of the harmonic potential, which 
will be chosen to be the $x$ direction.  For definiteness, the rotation of the 
interferometer will be in the $z$ direction and
the trap will be moved in the $y$ direction. 
An atomic cloud at temperature $T$ remains in equilibrium if the
center of the trap
is translated adiabatically. The trajectory of the moving trap
$y_0(t)$ is adiabatic when
\begin{equation}
\frac{d^2 y_0}{dt^2} \ll  \sqrt{T} \omega_\perp, \label{y0inequ}
\end{equation}
where $\omega_\perp$ is the trap frequency in the $y$ direction.

When Eq.~(\ref{y0inequ}) is fulled, the equations of motion 
Eq.~(\ref{lasersON1}) and (\ref{BoltzGeneralRot}) can be recast 
in a one-dimensional form.  The Wigner function is written as the
product $f(x,r_\perp, p, p_\perp) = f(x,p) F(r_\perp, p_\perp)$, where
$F(r_\perp, p_\perp)$ is the equilibrium distribution in the
perpendicular direction, normalized to one, and $f(x,p)$ is the non-equilibrium
distribution in the parallel direction, normalized to the number of
atoms in the trap.   
When the laser beams are on, the one-dimensional equation of motion
for the Wigner function is
\begin{equation}
\frac{\partial f}{\partial t} = \chi \sin(x) \left[
f(x, p-1/2) - f(x, p+1/2)
 \right], \label{eq:LasOn}
\end{equation}
The solution of Eq.~(\ref{eq:LasOn}) can be written in terms of Bessel
functions of the first kind $J_\nu$,
\begin{equation}
f(x,p,t) = \sum_{lk} (-i)^{l} J_k(\Xi) J_{l+k}(\Xi) e^{ i  (l + 2k )  x} f_0( x, p-l/2 ), \label{LaserSolution2}
\end{equation}
where the sum over $k$ and $l$ runs from $-\infty$ to $\infty$, $\Xi =
\int dt' \chi(t')$ is the strength of the laser pulse, and
\begin{equation}
  \label{eq:equilibDist}
  f_0 = \frac{N \omega_\parallel}{2 \pi T} \exp\left[ \frac{p^2 +
      \omega_\parallel^2 x^2}{2 T} \right] 
\end{equation} 
is the equilibrium Wigner function at temperature $T$.
In general Eq. (\ref{LaserSolution2}) can be negative, because the resulting gas is in a non-classical state.
However, for high temperatures $T \gg 1$ the negative parts of the
Wigner function are negligible, and the gas may be treated classically.

After the laser pulse is applied, the optical field is turned off and
the one-dimensional equation of motion for the Wigner function becomes

\begin{eqnarray}
 && \left[
\frac{\partial}{\partial t} + ( p - \Omega y_0) \frac{\partial}{\partial x} + \omega_\parallel^2 x \frac{\partial}{\partial p} 
 \right] f = \nu (f_0 - f), 
 \label{eq:DimensionlessEQM}
\end{eqnarray}
where the collision frequency $\nu$ is given by Eq.~(\ref{eq:nu}).
The left hand side of Eq.~(\ref{eq:DimensionlessEQM}) can be greatly simplified 
by introducing the new coordinates
\begin{eqnarray}
x' &=& x \cos \omega_\parallel t - \frac{p}{\omega_\parallel} \sin \omega_\parallel t
+ \Omega \int^t d \tau y_0(\tau) \cos \omega_\parallel \tau
\nonumber \\
p' &=& \omega_\parallel x \sin \omega_\parallel t + p \cos \omega_\parallel t
+ \Omega \omega_\parallel \int^t d \tau y_0(\tau) \sin \omega_\parallel \tau
\nonumber \\
t' &=& t. \label{eq:newCoodSys}
\end{eqnarray}
In this new coordinate system, Eq.~(\ref{eq:DimensionlessEQM}) becomes
\begin{equation}
 \frac{\partial}{\partial t} f = \nu ( f_0 - f), \label{BoltzmanNew}
\end{equation}
which has the general solution
\begin{equation}
 f(t) = f(0) e^{- \nu t} + f_0, \label{ZerothorderSolution}
\end{equation}
where $f(0)$ is the initial Wigner function given by Eq.~(\ref{LaserSolution2}), $f_0$ is the equilibrium 
Wigner function given by Eq.~(\ref{eq:equilibDist}), and $\nu$ is given by Eq.~(\ref{eq:nu}).

To read out the accumulated phase, the atomic cloud is illuminated with a single off resonate
laser beam.  The light that is back scattered off of the cloud is 
mixed with a reference beam \cite{cahn97, Wu07_2}.  By measuring the interference 
intensity, the amplitude of the scattered light can be determined.  
Using the Born approximation, it can be shown that the amplitude  
of the back scattered light is proportional to \cite{Wu07_2} 
\begin{equation}
 S = \int dx dp e^{i x} f(x,p),
\end{equation}
where $f$ is the one-dimensional Wigner function.  The quantity $S$ will 
be referred to as the interference signal of the
interferometer.

In the new coordinate system Eq.~(\ref{eq:newCoodSys}), the signal becomes
\begin{equation}
 S = e^{-i \varphi} \int dx dp
e^{i (x \cos \omega t + p/\omega \sin \omega t)} f \label{eq:newSignal}
\end{equation}
where
\begin{equation}
 \varphi = \Omega \left( \cos \omega t \int^t dt' y_0(t') \cos \omega t' + \sin \omega t 
\int^t dt' y_0(t') \sin \omega t'1
\right) \label{eq:phase_shift_general}
\end{equation}
is the phase shift due to the rotation of the interferometer.

Substituting Eq.~(\ref{ZerothorderSolution}) into Eq.~(\ref{eq:newSignal}) yields
\begin{eqnarray}
 S &=& N  e^{-i \varphi} \left\{
\sum_{nk} (-i)^n J_k J_{n+k} e^{
-\frac{T}{2 \omega^2} \left[ (\cos \omega t + n + 2k)^2 + \sin^2 \omega t \right]
+ i\frac{n}{2 \omega} \sin \omega t - \nu t 
} + e^{-T/2\omega^2}
\right\} . \label{SignalExact}
\end{eqnarray}
The interference signal $S$ is only nonzero when  $t \approx 2 n \pi /
\omega$, where $n$ is an integer or half integer.  
Expanding Eq.~(\ref{SignalExact}) near these points and taking the limit 
where $T \gg \omega^2$ yields 
\begin{equation}
 S = -2 N e^{-i\varphi - T  \tau^2 / 2  - \nu t} \sum_{k=0}^\infty J_k J_{k+1}
 \sin( (k+1) \tau / 2) \label{eq:SignalFinal} 
\end{equation}
where $\tau = t - 2 n \pi/\omega_\parallel$.
Equation~(\ref{eq:SignalFinal}) along with Eq.~(\ref{eq:nu}) and
(\ref{eq:phase_shift_general}) are the main analytical results of this paper.

\section{Discussion}\label{sec:discussion}

Equation (\ref{eq:SignalFinal}) will now be analyzed and optimal
operating conditions for trapped thermal atom interferometers will be found.  
Additionally to confirm the results of the
analytic model we will use a direct simulation Monte-Carlo model
(DSMC) of the interferometer \cite{Garcia00}.  

Direct simulation Monte-Carlo is accurate because the equation of motion after the
laser pulse Eq.~(\ref{BoltzGeneralRot}) is equivalent to the classical
Boltzmann equation.
Since the effect of the standing wave laser pulse on the cloud is non-classical
(Eq.~(\ref{eq:LasOn}) ), DSMC can only be used to model the dynamics
when the standing wave laser beams are off.  To account for the
laser pulse, the initial conditions for the DSMC model was set by
Eq.~(\ref{LaserSolution2}).   In regions where the initial Wigner function 
is negative $f < 0$, the classical distribution of atoms was set to
zero.  This is valid when the temperature is much larger than the
two-photon recoil temperature $T \gg 1$.

For definitiveness, we specialize to the case where the trap is moved back and fourth
according to 
\begin{equation}
\label{eq:y0}
 y_0 = \frac{d}{2} \cos(\omega_\parallel t),
\end{equation}
where $d$ is the dimensionless distance that the atomic cloud is
displaced in the $y$-direction.  Our chip will displace the trap about
$5~\mbox{mm}$, for $^{87}\mbox{Rb}$ the dimensionless displacement
will be $d = 8 \times 10^4$.

After $n/2$ oscillations, the atoms scattered into the first
order will enclose an area $\pi n d / \omega_\parallel$ and the accumulated
phase shift is
\begin{equation}
\label{eq:phaseshift}
 \varphi = \frac{\pi n d }{\omega_\parallel} \Omega.
\end{equation}
To measure the rotation rate of the Earth, with a $\pi$ phase shift in
a trap with $\omega_\parallel = 2 \pi \times 3~\mbox{Hz}$, the trap
must be moved back and forth three times.
The time that it takes for the interferometer to measure a given phase
shift does not depend on the parallel trap frequency.  For example the
time that it takes to measure a $\pi$ phase shift $t_\pi$ is
\begin{equation}
t_\pi = \frac{4 \pi}{\Omega d}.
\end{equation}
To measure Earth's rotation, with a $\pi$ phase shift, the
interferometer must have a cycle time of about one second.  To measure
a given rotation frequency, the bandwidth of the interferometer can
only be increased by increasing the distance that the atoms are
displaced $d$.  For the remainder of this paper only the interference
signal will be discussed for the case where the phase shift $\varphi$
is zero and when the trap is not moved in the $y$-direction.

\begin{figure}
\includegraphics[width=8.6cm]{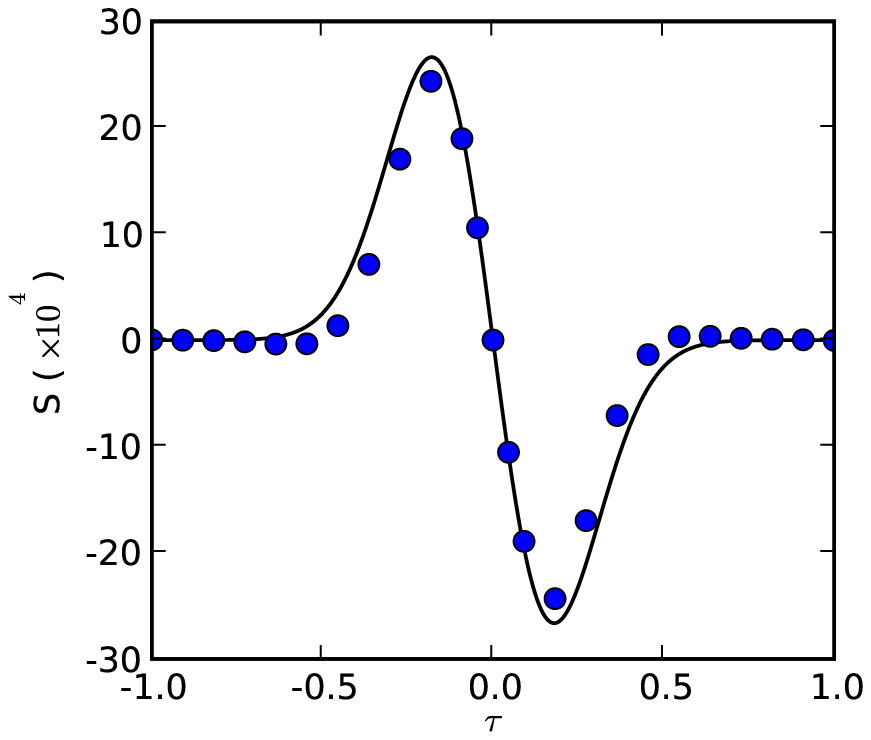}%
\caption{(color online) The interference signal as a function of
  time near one trap period.  The solid curve is
  Eq.~(\ref{eq:SignalFinal}) and the dots are the results of the DSMC
  code. \label{fig:Signal}}
\end{figure}
Figure~\ref{fig:Signal} shows the interferometer signal $S$, which is
proportional to amplitude of the back scattered
light, as a function of time.
The sold line in Fig. \ref{fig:Signal} is Eq.~(\ref{eq:SignalFinal}) for times close
to the first oscillation period $\tau = t - 2 n \pi / \omega_\parallel$
where $n=1$, and with the parameters $\sigma =0.2$, $\omega_\perp = 10^{-4}$,
 $\omega_\parallel = 10^{-2}$, $\varphi
= 0$, $N = 7
\times 10^6$, $\Xi = 1$ and $T = 30$.   The dots are the result of the DSMC, 
with each super particle representing 10 atoms and the signal averaged over $64$
separate runs of the DSMC code.  The error bars in all DSMC calculations are smaller than the
size of the dots shown in the figures.  This figure demonstrates good agreement between the analytic
result and our DSMC code.

The shape of this signal illustrates the time and position 
varying amplitude of the density modulation echo relative to the
probe laser.  For times slightly less than one trap period $\tau < 0$,
the nodes of the density modulation are located at the anti-nodes of
the standing wave laser field.  At precisely  one trap period, the
density modulation vanishes and the cloud returns to its initial
density distribution.  For times slightly larger than one trap period
$\tau > 0$, the nodes of the density modulation are located at the
nodes of the standing wave laser field.  

For weak pulses $\Xi \lesssim 1$, the interference signal 
Eq.~(\ref{eq:SignalFinal}) can approximately written as
\begin{equation}
S = - A \tau e^{-T\tau^2/2 - 2 n \nu \pi / \omega_\parallel}, 
\end{equation}
where $A = N \sum_k J_k J_{k+1} (k+1)$.  The two peaks in the signal
occurs at the times $\tau = \pm T^{-1/2}$.  For $^{87}\mbox{Rb}$ at
$40~\mu\mbox{K}$ the time between the maximum and minimum signal is
about $1~\mu\mbox{s}$.
The magnitude at the peaks in the signal is
\begin{equation}
 S_{peak} = A \frac{N}{T^{1/2}} \exp\left( - \frac{\omega_\perp^2 \sigma N n}{
   2^{1/2} \pi T} - \frac{1}{2}\right), \label{eq:SignalPeak}
\end{equation}
where Eq.~(\ref{eq:nu}) was used.  For the remainder of this section,
Eq.~(\ref{eq:SignalPeak}) will be analyzed for several illustrative
cases.  Using this analysis, limits on the performance of the
interferometer will be discussed.


\begin{figure}
\includegraphics[width=8.6cm]{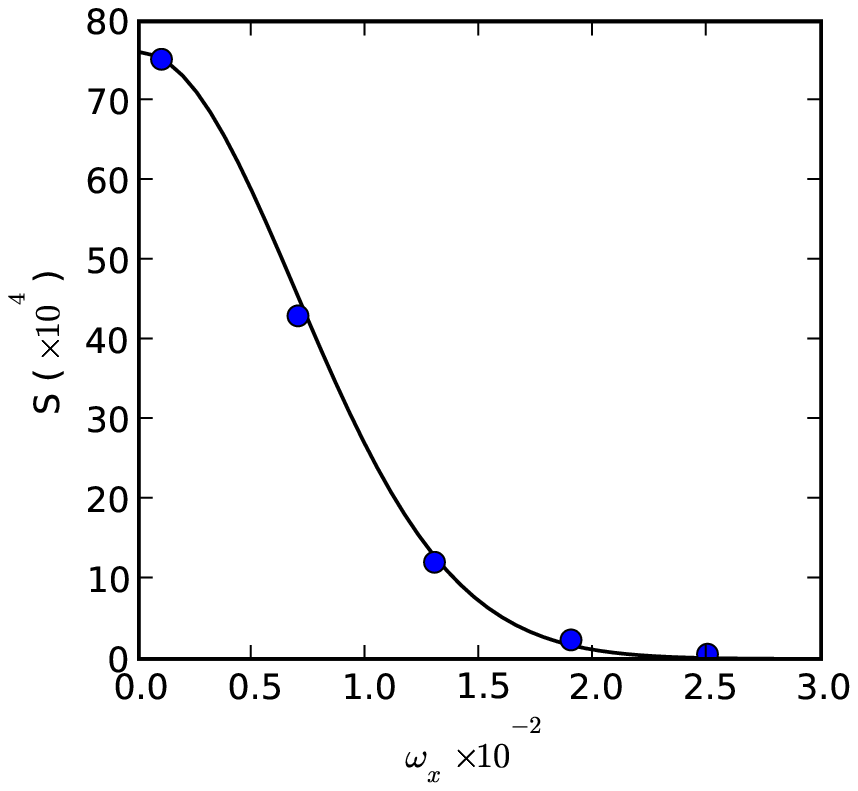}%
\caption{\label{fig:PeaksFreq} (color online) The maximum value of the interference
  signal as a function of perpendicular trapping frequency.  The solid
curve is found using Eq.~(\ref{eq:SignalPeak}) and the dots were
extracted from a DSMC calculation.}
\end{figure}
Figure \ref{fig:PeaksFreq} shows the maximum value of the interference
signal as a function of perpendicular trapping
frequency $\omega_\perp$,  where all the other parameters are the same 
as in Fig.~\ref{fig:Signal}.  The
solid curve is found using Eq.~(\ref{eq:SignalPeak}) and the dots were
extracted from the DSMC calculation.  This figure demonstrates
excellent agreement between the analytic and DSMC models of the
interferometer.  The maximum interference signal
is observed for small values of the perpendicular  trapping frequencies.
This is because the density of the atoms decreases as the atoms are
confined less tightly in the perpendicular direction.  

From Fig.~\ref{fig:PeaksFreq}, it is clear that the optimal value of
perpendicular trapping frequency is the smallest value such that the
movement of the trap remains adiabatic.  The minimum transverse
trap frequency can be estimated by using Eq.~(\ref{y0inequ}) and
(\ref{eq:y0}).  To remain adiabatic, the ratio
between the transverse and perpendicular trapping frequencies must be
\begin{equation}
  \label{eq:trapfreqratio}
  \frac{\omega_\perp}{\omega_\parallel} \gg \frac{\omega_\parallel
    d}{2 \sqrt{T}},
\end{equation}
where $d$ is the maximum displacement of the trap in the perpendicular 
direction and $T$ is the temperature of the gas.


\begin{figure}
\includegraphics[width=8.6cm]{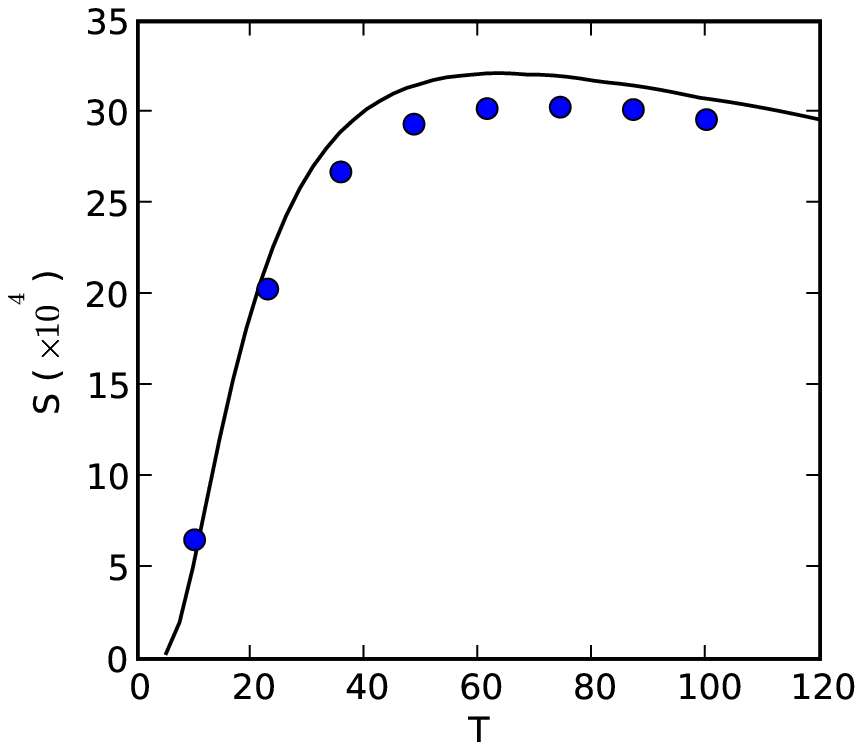}%
\caption{(color online) The maximum value of the interference signal
  as a function of temperature.  The solid curve is
  Eq.~(\ref{eq:SignalPeak}) and the dots were extracted from a DSMC
  calculation.  \label{fig:PeaksTemp}}
\end{figure}
Figure \ref{fig:PeaksTemp} shows the maximum value of the interference
signal as a function of the temperature $T$
of the trapped gas.  The remaining parameters are the same as
Fig.~\ref{fig:Signal}.  The solid curve is Eq.~(\ref{eq:SignalPeak})
and the dots were extracted from the DSMC calculation.  There is still
good agreement between the analytic and DSMC models.  

Holding all other parameters constant, the interference signal becomes
smaller as the temperature is reduced.  This is because, in a harmonic
potential, collision rate is inversely proportional to
temperature.   
For the parameters used in Fig.~\ref{fig:PeaksTemp}, the signal
increases with temperature until $T=60$.  For temperatures larger than
 $T>60$, the duration of the echo becomes shorter and the amplitude of the
density modulation is reduced.  
Using Eq.~(\ref{eq:SignalPeak}) it can be shown that the largest
amplitude of back-scattered light occurs when the temperature is
\begin{equation}
  \label{eq:optimalTemp}
  T = \frac{2^{1/2} \omega_\perp^2 \sigma N n}{\pi}.
\end{equation}
Equation (\ref{eq:optimalTemp}) shows that the optimal temperature
increases linearly with atom number.  As the atom number increases,
the signal to noise ratio of the detected signal decreases.  The 
time between the maximum and minimum amplitude decreases. The speed of
the detection scheme places an upper limit on the temperature and
therefore the lower limit on the signal to noise ratio.  Analysis of
the details of the detection scheme are beyond the scope of this
paper and will be left to future work.

The initial temperature of the atomic gas depends on the details of
the laser cooling and loading of the gas into the trap.  Although it
is possible to experimentally vary the final temperature, it is 
easier to  vary the number of trapped atoms.  This can be done
by changing the load time of the magneto optical trap.  Because of this, we believe
that it is most useful to treat temperature $T$, and trap frequencies $\omega_\parallel$ and
$\omega_\perp$ as constants and optimize the number of trapped atoms $N$.


\begin{figure}
\includegraphics[width=8.6cm]{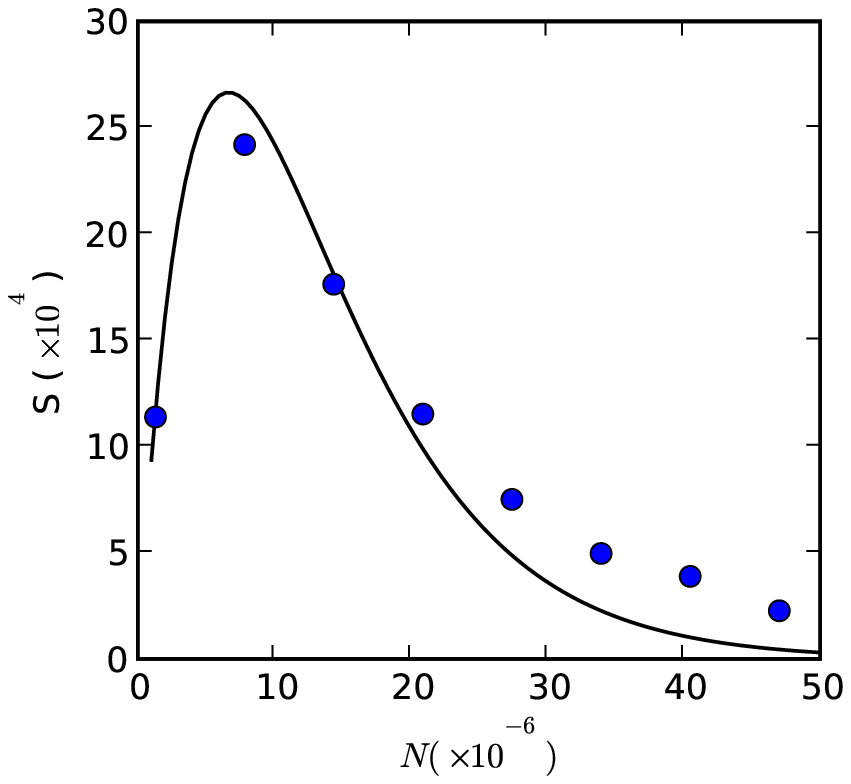}%
\caption{(color online) The maximum value of the interference signal
  as a function of number of atoms in the trap.  
The solid curve is Eq.~(\ref{eq:SignalPeak}) and the dots were
extracted from a DSMC calculation.
\label{fig:PeaksNumber}}
\end{figure}

Figure \ref{fig:PeaksNumber} shows the maximum value of the
interference as a function of number of trapped atoms $N$.  The solid curve was found using
Eq.~(\ref{eq:SignalPeak}) and the dots were extracted from our DSMC
code.  The remaining parameters are the same as in
Fig.~\ref{fig:Signal}.  When the number of atoms in the gas is $N  < 7
\times 10^6$, the signal increases 
with increasing number.  This is because as the number of atoms
increases so does the amount of back scattered light.  When the
total number of atoms in the gas is $N > 7 \times 10^6$ the interference signal
decreases because the higher density increases the collision rate
between the atoms in the gas.
Using Eq.~(\ref{eq:SignalPeak}) it can be shown that, holding all other
parameters constant, the maximum value
of the interference signal occurs the number of atoms is
\begin{equation}
\label{eq:optimalNumber}
  N =  \frac{2^{1/2} \pi T}{n \omega_\perp^2 \sigma}.
\end{equation}
For a trap with frequencies $\omega_\parallel = 2 \pi \times 3
\mbox{Hz}$ and $\omega_\perp = 2 \pi \times 300 \mbox{Hz}$, that traps
$^{87}\mbox{Rb}$ atoms at $40~\mu\mbox{K}$ the optimal number of atoms
for one trap period is about $7 \times 10^{6}\mbox{atoms}$.  To
measure Earth's rotation with at $\pi$ phase shift, by displacing the trap by
$5~\mbox{mm}$, the atoms must oscillate three time in this trap and
the optimal number of atoms is $2.2 \times 10^{6}$.

\section{Conclusions}\label{sec:conclusions}

In this paper we presented a simple analytic model of the dynamics of
a trapped atom interferometer that uses a single Kapitza-Dirac pulse to
modulate the atoms and the classical turning points of the trap to
reflect them.  The interferometers signal is read out by reflecting a
single probe pulse off of the atoms and interfering the back-reflected
light with a reference beam.  We presented
a description of the collisions between the atoms and showed
that our simple model give quantitatively accurate results when compared
to a DSMC model of the interferometer.  Finally, we used our model to
find the optimal temperature or number to maximize the performance of
the interferometer.  

Although the analytic model presented in this paper specialized to the
analysis of a single Kapitza-Dirac pulse, the results of
Sec.~\ref{sec:discussion} easily generalize to multi-pulse
interferometers \cite{Wu07_2, su08}.  To apply our model to interferometers that use gases above the BEC
transition temperature but below the recoil temperature the momentum and spacial dependence on the collision
frequency cannot be ignored and Eq.~(\ref{nu1}) must be used instead
of Eq.~(\ref{eq:nu}).  We
believe that inclusion of the more complicated collision frequency
will not dramatically change the results of this paper when describing
a gas below the recoil temperature.

\section{Acknowledgements}

We would like to thank Alexey Tonyushkin, Mara Prentiss, Alex Zozulya,
and Val Bykovsky for their useful conversations on this subject. 
The authors acknowledge support from the Air Force Office of
Scientific Research under program/task 2301DS/03VS02COR and DARPA.



\end{document}